\documentclass[preprint,showpacs,aps,preprintnumbers,amsmath,amssymb]{revtex4-1}

\usepackage{graphicx,epsfig}
\usepackage{dcolumn}
\usepackage{bm}
\newcommand{\be}{\begin{equation}}
\newcommand{\ee}{\end{equation}}
\newcommand{\bse}{\begin{subequations}}
\newcommand{\ese}{\end{subequations}}
\newcommand{\bary}{\begin{eqnarray}}
\newcommand{\eary}{\end{eqnarray}}
\newcommand{\bwt}{\begin{widetext}}
\newcommand{\ewt}{\end{widetext}}

\begin{document}


\title{The origin of multi-TeV flares from the nearest blazar\\ Markarian 421}
\author{Sarira Sahu$^{a,b}$ }
\email{sarira@nucleares.unam.mx}
\author{Alberto Rosales de Le\'{o}n$^{a}$}
\email{albertoros4@ciencias.unam.mx}
\author{Shigehiro Nagataki$^{b,c,d}$}
\email{shigehiro.nagataki@riken.jp}
\author{Virendra Gupta$^{e}$}
\affiliation{$^{a}$Instituto de Ciencias Nucleares, Universidad Nacional Aut\'onoma de M\'exico, 
Circuito Exterior, C.U., A. Postal 70-543, 04510 Mexico DF, Mexico}


\affiliation{$^{b}$Astrophysical Big Bang Laboratory, RIKEN, Hirosawa, Wako, Saitama 351-0198, Japan}
\affiliation{$^{c}$Interdisciplinary Theoretical Science Research
  (iTHES), RIKEN, Hirosawa, Wako, Saitama 351-0198, Japan}

\affiliation{$^{d}$Interdisciplinary Theoretical \& Mathematical Science (iTHEMS),
  RIKEN, Hirosawa, Wako, Saitama 351-0198, Japan}
\affiliation{$^{e}$Departamento de  F\'isica Aplicada, 
Centro de Investigac\'ion y de Estudios Avanzandos del IPN\\
Unidad M\' erida, A. P. 73, Cordemex, M\'erida, Yucat\'an, 97310, Mexico
}

\begin{abstract}

Markarian 421 is a high-peaked BL Lac object and it has undergone many
strong outbursts since its discovery as a TeV source in
1992. Markarian 421 has been stud- ied intensively and was observed by
various Cherenkov tele- scope arrays ever since. The outbursts of
April 2004 observed by the Whipple telescope and of February 2010 by
the HESS telescopes are explained well in this work by using the
photohadronic model. To account for the attenuation of these high-
energy gamma-rays by the extragalactic background light (EBL), we use
template EBL models. The intrinsic spectrum of each epoch is different
even though the high-energy protons have almost the same spectral
index. We observe that this difference in intrinsic spectra is due to
the change in the spectral index of the low-energy tail of the
synchrotron self Compton (SSC) photons during different epochs of
flaring. Our results show that the contemporaneous multiwavelength
observations, particularly in the low-energy tail region of the SSC
emission of the source, are important in explaining the flaring
phenomenon.

\end{abstract}

\maketitle

\section{Introduction} \label{sec:intro}

The extragalactic very high energy (VHE, $E_{\gamma} > 100$ GeV) gamma-rays undergo
energy dependent attenuation en route to
Earth by the intervening extragalactic background light (EBL)
through electron-positron pair production\cite{Stecker:1992wi}. This interaction process
not only attenuates the absolute flux but also significantly 
changes the VHE emission spectrum. 
The diffuse EBL contains a record of the star
formation history of the Universe. A proper 
understanding of the EBL
spectral energy distribution (SED) 
is very important for the correct interpretation of the
deabsorbed VHE spectrum from the source.
The direct measurement of the EBL is very difficult with
high uncertainties mainly due to the contribution of
zodiacal light\cite{Hauser:2001xs,Chary:2010dc}, and
galaxy counts result in a lower limit since the number of unresolved
sources (faint galaxies) is unknown\cite{Madau-Po:2000}. 
Keeping in mind
the observational constraints, several approaches with different
degrees of complexity have been developed to
calculate the EBL density as a function of energy for different
redshifts\cite{Dominguez:2010bv,Inoue:2012bk,Franceschini:2008tp, Kneiske:2002wi,Stecker:2005qs,Stecker:2016fsg,Orr:2011ha,Primack:2005rf}. Mainly three types of EBL models exist: backward and forward evolution models
and semi-analytical galaxy formation models with a combination of
information about galaxy evolution and observed properties of galaxy spectra.
In the backward evolution models\cite{Stecker:2005qs},  one starts from the observed
properties of galaxies in the local universe and evolve them from cosmological initial conditions or
extrapolating backward in time using parametric models of the
evolution of galaxies. This extrapolation induces
uncertainties  in the properties of the EBL which increases at high redshifts.
On the contrary, the forward evolution
models\cite{Franceschini:2008tp,Kneiske:2002wi} predict the temporal
evolution of 
galaxies forward in time starting from the cosmological
initial conditions. 
Finally, semi-analytical models have been developed
which follow the formation of large scale structures driven by cold dark
matter in the universe by using the cosmological parameters from
observations. This method also accounts for the merging of the dark matter
halos and the emergence of galaxies which form as baryonic matter falls into the
potential wells of these halos.


Blazars are a
subclass of AGN and the dominant extra galactic population in
$\gamma$-rays\cite{Acciari:2010aa}. 
These objects show rapid variability in the entire
electromagnetic spectrum and have non-thermal spectra produced by the relativistic jet of plasma oriented close
to the observers line of
sight\cite{Urry:1995mg}. The jets are powered by matter accretion onto
supermassive black hole. 
The spectral energy distribution (SED) of these blazars
has a double peak structure in the $\nu-\nu F_{\nu}$ plane. 
The low  energy peak corresponds to
the synchrotron radiation from a population of relativistic electrons
in the jet  and the high energy
peak believed to be due to the synchrotron self
Compton (SSC) scattering of the high energy electrons with their
self-produced synchrotron photons
\cite{Dermer:1993cz,Sikora:1994zb}. As leptons ($e^{\pm}$) are responsible for the
production of the SED, this is called the leptonic model and in general is
very successful in explaining the multiwavelength emission from blazars and FR~I galaxies\cite{Fossati:1998zn,Ghisellini:1998it,Abdo:2010fk,Roustazadeh:2011zz}.

Blazars detected in
VHE are predominantly high energy peaked blazars (HBLs) whose
synchrotron peak lies in the UV to X-ray energy range and the inverse Compton peak
is in the GeV-TeV energy range\cite{Fossati:1998zn,Urry:1995mg}.
Flaring seems to be a major activity of these blazars
which is unpredictable and switches between quiescent and
active states involving different time scales and flux
variabilities\cite{Albert:2007zd,Aharonian:2007ig,Ghisellini:2008gn,Ghisellini:2008us,Resconi:2009sw}. 
However, broadly speaking the flaring mechanism is unknown.


From the continuous monitoring and dedicated multi-wavelength
observations of the nearest HBLs Markarian 421 (Mrk 421, z=0.0308\cite{Gorham:1999en,Ulrich:1975ApJ}), Mrk 501 (z=0.033 \cite{Ulrich:1975ApJ,Stickel:1993aap}) and 1ES 1959+650 (z=0.047\cite{Schanchter:1993ApJ}), several major multi-TeV flares have been observed\cite{Tluczykont:2011gs,Aharonian:2000nr,Aliu:2016kzx,Chandra:2017vkw,Santander:2017wjl}. 
Strong temporal correlation in different wavebands, particularly in
X-rays and VHE $\gamma$-rays  has been observed in some flaring
events, however, in some other flaring events no such correlation is
observed \cite{Krawczynski:2003fq,Blazejowski:2005ih}, which seems unusual for a leptonic origin\cite{Katarzynski:2006db,Fossati:1998zn,Ghisellini:1998it,Roustazadeh:2011zz} of the multi-TeV
emissions and needs to be addressed through other alternative
mechanisms\cite{Mucke:1998mk,Mucke:2000rn,Cao:2014nia,Zdziarski:2015rsa,Essey:2010er,Ghisellini:2004ec,Tavecchio:2008be}. 

Mrk 421 is the first extragalactic source detected in the multi-TeV domain\cite{Punch:1992xw} and it is one of the fastest varying $\gamma$-ray
sources. There are other BL Lac objects with lower redshifts than Mrk 421, however, these objects were never reported as TeV emitters and could have been misclassified\cite{VeronCetty:2006zz}. Also, there are many BL Lacs with unknown redshifts.

Several large flares of Mrk 421 were observed in 2000 -
2001\cite{Okumura:2002jw,Amenomori:2003dy,Fossati:2007sj} and 
2003 - 2004\cite{Cui:2004wi,Blazejowski:2005ih}. During April 2004, 
large flare took place both in the X-rays and in the 
TeV energy band. The source was observed
simultaneously at TeV energies with the Whipple 10 m telescope and at
X-ray energies with the Rossi X-ray Timing Explorer (RXTE)\cite{Blazejowski:2005ih}. It was
also observed simultaneously in radio and optical wavelengths. 
During the flaring period, the TeV flares had no
coincident counterparts at longer wavelengths and 
it was observed that the X-ray flux reached its peak 1.5 days before the
TeV flux did during this outburst. 
The orphan TeV flare in 1ES 1959+650 of 2002\cite{Krawczynski:2003fq}
had also no simultaneous X-ray counterpart and the variation
pattern in X-rays were similar to the one observed in Mrk 421.
A strong outburst in multi-TeV
energy in Mrk 421 was first detected by VERITAS telescopes on 16th of February
2010 and follow up observations were done by the HESS telescopes
during four subsequent nights\cite{Tluczykont:2011gs}.

In the framework of a six month long multi-instrument campaign, the
MAGIC telescopes observed VHE flaring from Mrk 421 on 25th of
April 2014 and the flux (above 300 GeV) was about 16 times brighter
than the usual one. This triggered a joint ToO program by XMM-Newton,
VERITAS, and MAGIC. These three instruments individually observed
approximately 3 h each day on April 29, May 1, and May
3 of 2014\cite{Abeysekara:2016qwu}. The simultaneous VERITAS-XMM-Newton observation is
published recently and it is shown that the observed multiwavelength spectra are
consistent with one-zone synchrotron self-Compton model\cite{Abeysekara:2016qwu}. However, 
the details of the large flare observed on 25th April by MAGIC are not yet publicly available.

Mrk 421 being the nearest HBL, the VHE gamma-rays can be attenuated
the least and in principle presently operating Cherenkov telescopes
will be able to observe higher energy photons. At the same time,
different EBL models can also be compared.
So Mrk 421 can be used as an example to study the intrinsic emission
mechanism in the multi-TeV energy range from the nearest blazars and test different EBL models.

From the above various VHE emission epochs, we shall analyze the
following events:
(1) the flare of April 2004\cite{Blazejowski:2005ih} and (2) the flare
of February 2010\cite{Balokovic:2015dnz}.
Both the experimental data sets are public and have
multiwavelength information, which is important to improve the
photohadronic fitting of the flare. Also both flares were having
high flux which is important for our study. The flare of 2004 was
analyzed by Sahu et al.\cite{Sahu:2015tua} where EBL
correction was not considered. In this work we include the EBL correction for the same 2004 data.
For our analysis of these
flaring events, we will be using the photohadronic model, 
a detailed account of this model is given in refs. \cite{Sahu:2012wv,Sahu:2013ixa,Sahu:2013cja,Sahu:2015tua,Sahu:2016mww,Sahu:2016bdu,Sahu:2018rqs}.
This model is successfully employed to explain the multi-TeV flaring from many
high frequency blazars. A brief account of the model is given in the
next section.

\section{Photohadronic Scenario}
In the photohadronic scenario, the Fermi accelerated protons having a power-law spectrum $dN/dE_p \propto E^{-\alpha}_p$ interact with the background photons to produce the $\Delta$-resonance which subsequently decays to gamma-rays via intermediate neutral pion and to neutrinos through charged pion\cite{Sahu:2012wv}. We assume that this interaction occurs within a compact and confined volume of radius $R'_f$ inside the blob of radius $R'_b$ ($R'_f < R'_b$), here the $^{\prime}$ notation implies the jet comoving frame. Geometrically this represents a double jet structure, one compact and smaller cone which is enclosed by a bigger one along the same axis, the geometry of this model is discussed in Fig. 1 of ref. \cite{Sahu:2015tua}. The inner compact region has a photon density much higher than the
outer region. Due to the adiabatic expansion of the inner jet, the 
photon density will decrease when it crosses into the outer
region. We assume the scaling behavior of the photon densities in
the inner and the outer jet regions which essentially means, the spectra of both the
outer and the inner jets have the same slope. Mathematically we can express this as
\be
\frac{n'_{\gamma, f}(\epsilon_{\gamma_1})}
{n'_{\gamma, f}(\epsilon_{\gamma_2})} \simeq \frac{n'_\gamma(\epsilon_{\gamma_1})}
{n'_\gamma(\epsilon_{\gamma_2})},
\label{densityratio}
\ee 
i.e. the ratio of photon densities at two different
background energies $\epsilon_{\gamma_1} $  and $\epsilon_{\gamma_2} $
in the flaring ($n'_{\gamma, f}$) and in the non-flaring ($n'_{\gamma}$)
states remains almost the same. 
The photon density $n'_{\gamma}$ in the outer region can be calculated in the usual
way using the observed/fitted SED. Afterwards, by using the above
relation in Eq. (\ref{densityratio}), we can express
the unknown inner photon density in terms of the outer known
density. Henceforth, for our calculation, we shall use $n'_{\gamma}$ and its corresponding
flux rather than the one from the inner jet region.

The observed VHE $\gamma$-ray flux depends on the background seed
photon density and the differential power-spectrum of the Fermi
accelerated protons given as
$F_{\gamma}\propto n'_{\gamma} (E^2_p\,dN/dE_p)$.
It is to be noted that, the photohadronic process in a standard
blazar jet environment (quiescent state) is inefficient due to low seed photon
density $n'_{\gamma}$. So to explain the multi-TeV emission from the
flaring in the photohadronic scenario, 
jet kinetic power has to be increased to the super-Eddington limit\cite{Cao:2014nia,Zdziarski:2015rsa}.
However, the inner compact jet scenario evade this problem due to the
higher photon density\cite{Sahu:2013ixa}. 
So, the assumption here is that the outer jet is always there and is responsible
for the quiescent state of the blazar while the inner jet is transient and
is responsible for the flare.
The photohadronic model provides a simple explanation for the
multi-TeV observed events from the HBLs and it depends only on the
seed photons from the low-energy tail of the SSC emission. On the other hand, the one zone
leptonic model does not fit well to the observed multi-TeV data and the
multi-zone leptonic model needs many more free parameters (compared to one zone leptonic model) to fit it.

The observed $\gamma$-ray energy $E_{\gamma}$ from the $\pi^0$ decay
and the background seed photon energy $\epsilon_{\gamma}$ satisfy the
kinematical condition
\be
E_{\gamma}  \epsilon_{\gamma} \simeq 0.032 \frac{{\cal
    D}^2}{(1+z)^2}\, {\rm GeV^2},
\label{KinCon}
\ee 
following the resonance process $p\gamma\rightarrow \Delta$ and the
incident proton has the energy $E_{p} \simeq 10E_{\gamma}$. 
Here ${\cal D}$ is the Doppler factor and the bulk
Lorentz factor of the blob is given as $\Gamma$. For blazars we have
$\Gamma\simeq {\cal D}$.

The different epochs of VHE flaring between 2004 to 2014 of Mrk 421
had different ranges of observed $\gamma$-rays which correspond to different
ranges of seed photon energies. 
We observed that, irrespective of the $E_{\gamma}$ and epoch of
flaring, the ranges of $\epsilon_{\gamma}$ are always in the low
energy tail region of the SSC emission. This is the region in the valley where the SSC emission begins (see red line in Fig. \ref{fig:figure2} where $0.26 {\rm MeV} \le E_\gamma \le 100 {\rm MeV}$).

\begin{figure}
{\centering
\resizebox*{0.5\textwidth}{0.35\textheight}
{\includegraphics{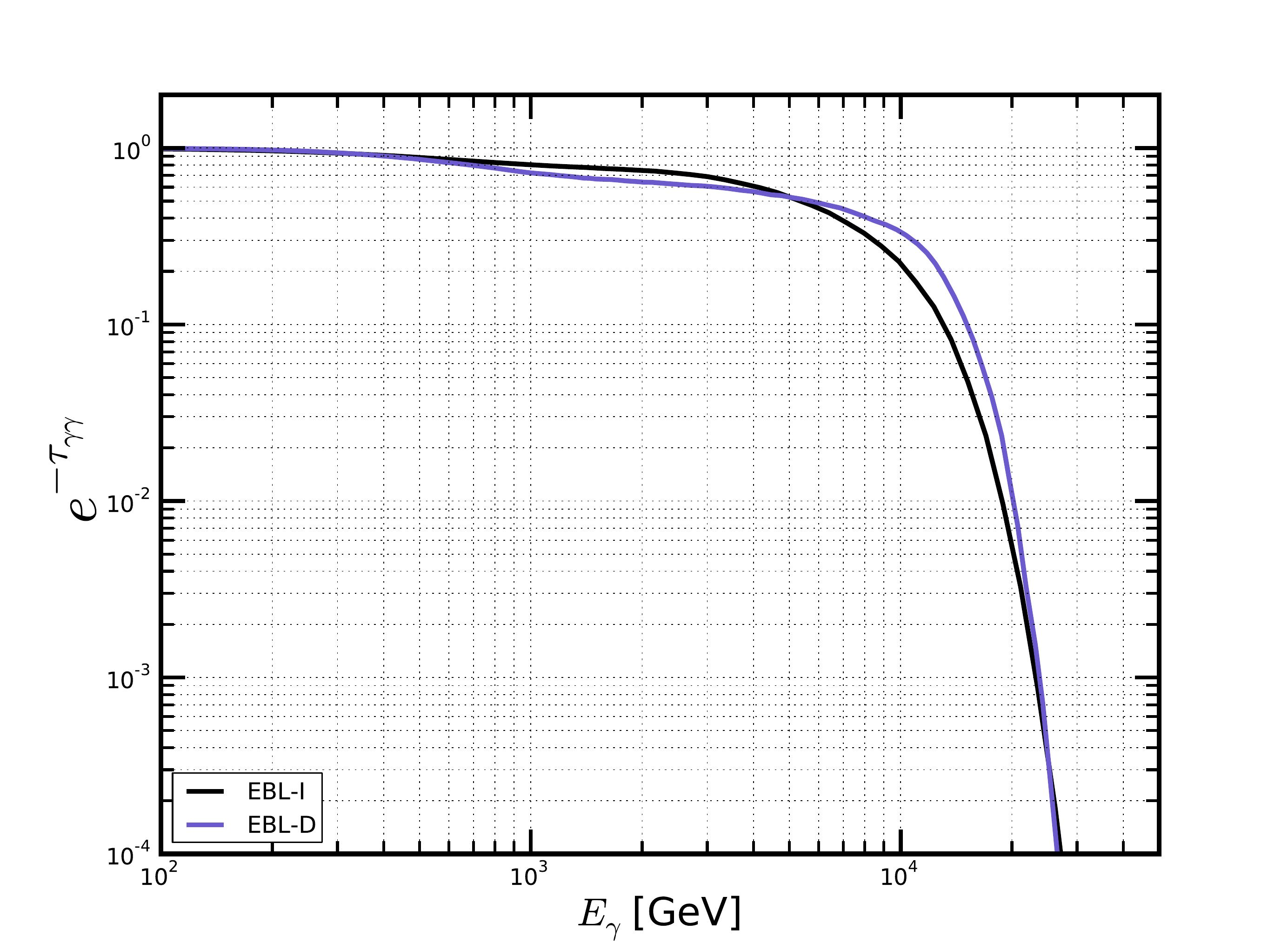}}
\par}
\caption{ 
The attenuation factor as a function of photon energy predicted by the two EBL models for $z=0.031$.
\label{fig:figure1}
}
\end{figure}

\begin{figure}
{\centering
\resizebox*{0.5\textwidth}{0.35\textheight}
{\includegraphics{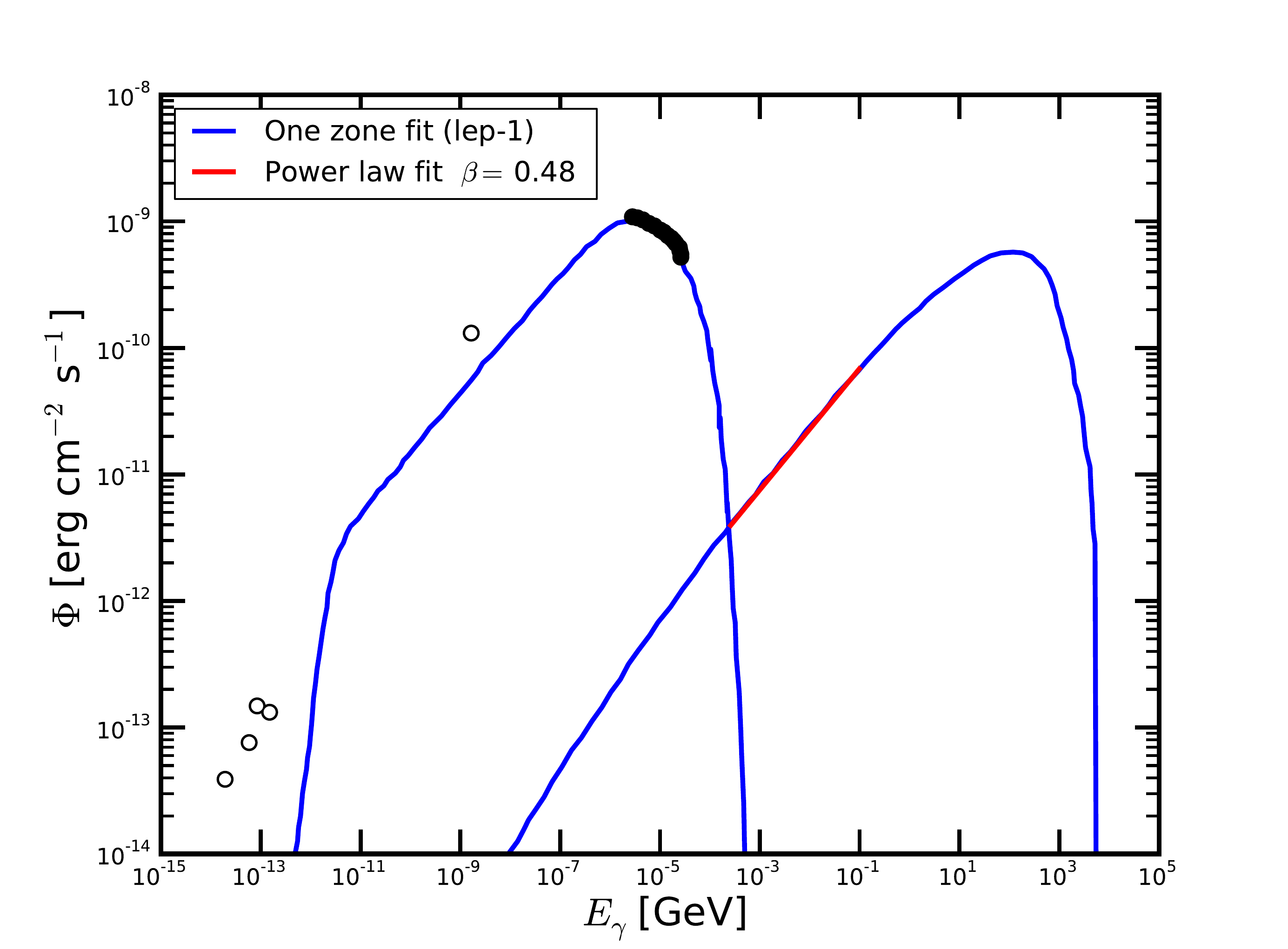}}
\par}
\caption{
The fit to the tail region of the SSC SED (lep-1)\cite{Blazejowski:2005ih} with the power-law
as given in Eq. (\ref{phi_power_law}) with $\Phi_0=6.0\times
10^{-10}\, erg\, cm^{-2}\, s^{-1}$ and $\beta=0.48$ (red curve).
\label{fig:figure2}
}
\end{figure}

The SSC flux in this range of seed
photon energy is exactly a
power-law given by $\Phi_{SSC} \propto \epsilon^{\beta}_{\gamma}$
with $\beta > 0$. Again, from the kinematical condition to produce
$\Delta$-resonance through $p\gamma$ interaction, $\epsilon_{\gamma}$
can be expressed in terms of $E_{\gamma}$ and can be written as
\be
\Phi_{SSC}(\epsilon_{\gamma}) =\Phi_0\, E^{-\beta}_{\gamma}.
\label{phi_power_law}
\ee
From the leptonic model fit to the observed multiwavelength data (up
to second peak) during a quiescent/flaring state we can get the SED
for the SSC region from which $\Phi_0$ and $\beta$ can be obtained easily. 
By expressing the observed flux $F_{\gamma}$ in terms of the
intrinsic flux $F_{\gamma,in}$ and the EBL correction as
\be
F_{\gamma} (E_{\gamma}) = F_{\gamma, in}(E_{\gamma})\,e^{-\tau_{\gamma\gamma}(E_{\gamma},z)}, 
\ee
where the intrinsic flux is given in ref. \cite{Sahu:2016bdu} as,
\be
F_{\gamma, in}(E_{\gamma}) =A_{\gamma}\,\Phi_0 \left (\frac{E_{\gamma}}{\rm TeV}
\right )^{-\alpha-\beta+3},
\ee
where $A_{\gamma}$ is a dimensionless normalization constant and can be fixed
by fitting the observed VHE data. As discussed above, the power
index $\beta$ is fixed from the tail region of the SSC SED for a given
leptonic model which fits the low energy data well. So the Fermi
accelerated proton spectral index $\alpha$ is
the only free parameter to fit the intrinsic spectrum. 
To account for the EBL correction to the observed multi-TeV gamma-rays, here we
use the EBL models of Dominguez et al.\cite{Dominguez:2010bv} (EBL-D) and Inoue et al.\cite{Inoue:2012bk} (EBL-I) to interpret our
results. The Fig. \ref{fig:figure1} presents the attenuation 
factor for these two models as functions of observed gamma-ray energy
$E_{\gamma}$ for $z=0.031$. 
The Fig. \ref{fig:figure1} shows that the difference between the two
models becomes apparent at 600 GeV, but continues to increase above 1
TeV,  the difference is maximum around 2 TeV than decreases until the
models intersect at 6 TeV. At higher energies the models diverge and
converge again above 20 TeV with the same attenuation factor. 

The Fermi accelerated protons in the jet will emit synchrotron
radiation but it will be suppressed by a factor of $m^{-4}_p$, where
$m_p$ is the proton mass. Also
the emission from the ultra high energy protons needs a stronger
magnetic field. The same ultra high energy protons can leak out from
the jet region and can reach to the Earth as ultra high energy cosmic
rays (UHECRs). The charged pions produced from the photohadronic
process will decay to neutrinos which can in principle be detected on
Earth\cite{Winter:2013cla,Padovani:2014bha,Krauss:2014tna}, however,
the flux is too low.

\section{Two flaring episodes}
As discussed in the introduction, several epochs of major multi-TeV
emission/flaring are observed from Mrk 421 and below we analyze 
two of these multi-TeV flares observed by different Cherenkov telescope arrays and use photohadronic model to interpret these flaring events. As input to the
photohadronic model we use the leptonic SSC SED fitted to the multiwavelength observations.

\begin{figure}
{\centering
\resizebox*{0.5\textwidth}{0.35\textheight}
{\includegraphics{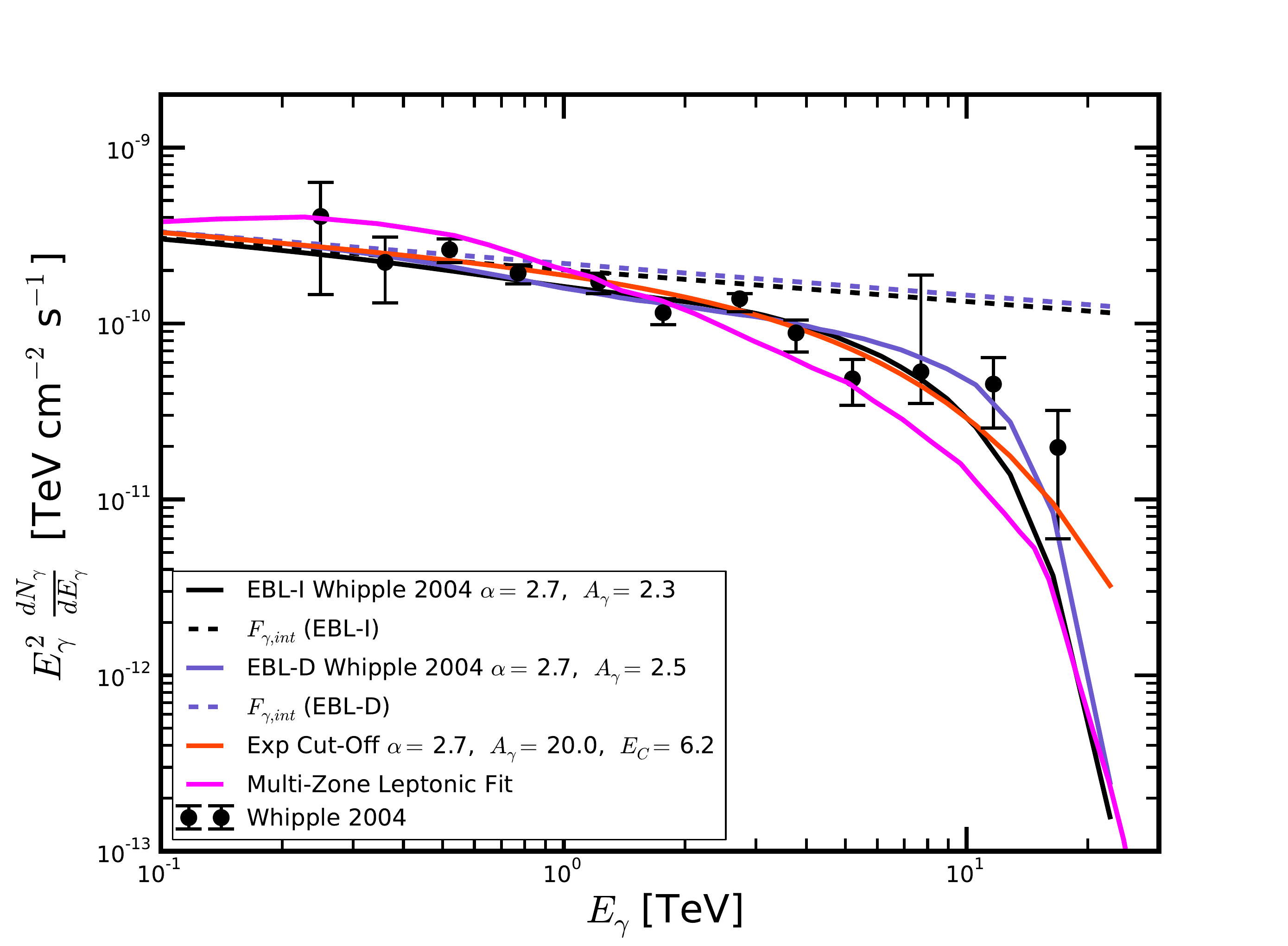}}
\par}
\caption{
Fit to the observed flux of April 2004 flare  with the photohadronic model
using two different EBL models are shown. It is also compared with the power-law
with exponential cut-off without EBL correction fit\cite{Sahu:2015tua} and with the
multi-zone leptonic fit\cite{Blazejowski:2005ih}. The multi-zone
leptonic model accounts for the attenuation of the very high energy 
gamma-rays by  the diffuse infrared background.The intrinsic fluxes for
both the EBL models are also shown.
\label{fig:figure3}
}
\end{figure}

\subsection{ The flare of April 2004}
The multi-TeV flare of April 2004 was the first flare observed in
multiwavelength and also it was difficult to explain by one-zone
leptonic model\cite{Blazejowski:2005ih}. The Whipple telescope
observed the flare in the energy range $ 0.25\, {\rm TeV} (6.0\times 10^{25}\, {\rm Hz}) \le E_{\gamma} \le 16.85\, {\rm TeV}  (4.1\times 10^{27}\, {\rm Hz}) $. We use the one-zone leptonic model of
ref. \cite{Blazejowski:2005ih} (lep-1) as input for the
photohadronic model to explain the observed TeV emission. Here the bulk Lorentz
factor associated with the lep-1 model is $\Gamma={\cal D}=14$. The
Fermi accelerated proton energy  lies in the range $2.5\, {\rm TeV} \le
E_p \le 168\, {\rm TeV}$ and the corresponding background photon energy is  
in the range $23.6\, {\rm MeV} (5.7\times 10^{21}\, {\rm Hz})$ $\ge$
 $\epsilon_{\gamma} \ge 0.35\, {\rm MeV}$ $ (8.4\times 10^{19}\, {\rm Hz}) $. This
range of $\epsilon_{\gamma}$ is in the low energy tail 
region of the SSC SED and the corresponding  flux follows an exact
power-law given in Eq. (\ref{phi_power_law}) 
with $\Phi_0 = 6.0\times 10^{-10}$ $ {\rm erg\, cm^{-2}\, s^{-1} }$ and
$\beta=0.48$ which is shown in Fig. \ref{fig:figure2}.

For the EBL correction to the observed data, we use EBL-D and EBL-I. 
The best fit to the observed multi-TeV flare data is obtained for  the spectral index
$\alpha=2.7$, and the
normalization constant $A_{\gamma}$ for EBL-D is  2.3 and for EBL-I
is 2.5 respectively which are shown in
Fig. \ref{fig:figure3}. For $E_\gamma < 4 TeV$
both these EBL models fit the data very well, and, above this energy 
there is a slight difference due to the change in the attenuation factor.
We observe that EBL-D fit is better than the EBL-I. Again above 20 TeV
both the EBL models behave the same. For comparison we have also
shown in the same figure (red curve) the photohadronic
model fit without EBL correction but with an exponential
cut-off\cite{Sahu:2015tua}, and the multi-zone leptonic model fit
(magenta curve)\cite{Blazejowski:2005ih}. It can be seen from Fig. \ref{fig:figure3} that the
multi-zone fit is not so good compared to other fits for $E_{\gamma}
\leq 15$ TeV. However, for higher energy it has the same behavior as EBL-D
and EBL-I.
 With the exponential cut-off scenario, 
a good fit is obtained for the spectral index $\alpha=2.7$ and the cut-off
energy $E_c=6.2$ TeV. Again comparing this with the EBL corrected models,
below 4 TeV,  all these
fits are exactly the same. However, above 4 TeV we observe some 
discrepancy among these fits and above 10 TeV the fits of EBL-D
and EBL-I fall faster than the exponential cut-off
scenario. It is clear from the comparison of EBL-I with the
exponential cut-off  scenario that, for $E_{\gamma}\leq 10$ TeV both
$e^{-E_{\gamma}/E_c}$ and $e^{-\tau_{\gamma\gamma}}$  are almost the same and
above 10 TeV the attenuation factor falls faster than the
exponential cut-off for which the EBL-D and EBL-I curves fall
fast. 

The intrinsic flux in EBL-D and EBL-I are also shown. Both are
almost the same and having power-law behavior with $F_{\gamma,
  in}\propto E^{-0.18}_{\gamma}$. Even though  all these models fit
quite well to the observed data below 20 TeV energy range, the deviation is
appreciable above 20 TeV between the EBL corrected plots and the exponential
cut-off. So observation of VHE flux above $\sim 30$
TeV will be a good test to constrain the EBL effect on the VHE
gamma-rays from Mrk 421 and to see whether energy cut-off scenario is necessary or not.

In ref.\cite{Sahu:2015tua}, by comparing expansion time scale,
interaction time scale of $p\gamma$ interaction and the high energy
proton luminosity to be smaller than the Eddington luminosity in the
inner jet region $R'_f\simeq 3\times 10^{15}\, {\rm cm}$ (size of the outer
jet is $R'_b\simeq 0.7\times 10^{16}\, {\rm cm}$), the range of
optical depth for the $\Delta$-resonance production is estimated as 
$0.02 \, < \tau_{p\gamma} \, < 0.13$. This corresponds to a photon
density in the inner jet region as $1.3\times 10^{10}\, {\rm cm^{-3}} <
n'_{\gamma, f} < 8.9\times 10^{10}\, {\rm cm^{-3}}$. 
The TeV photons produced from the neutral pion decay will mostly
encounter the SSC photons in the energy range $0.35\, MeV \leq
\epsilon_{\gamma} \leq 23.6\, MeV$. The pair production cross section
for $\epsilon_{\gamma} \geq 0.35\, MeV$ is very small
($\sigma_{\gamma\gamma} \leq 10^{-30}\, cm^{-2}$) which corresponds to a mean free path of
$\lambda_{\gamma\gamma} \geq 10^{19}\, cm$ for the
multi-TeV gamma-rays, larger than the outer jet size. So, the TeV
photons will not be attenuated much due to the $e^+e^-$ pair
production. The parameters used in
the photohadronic to fit the 2004 data are summarized in Table \ref{tab1}.

\subsection{The flare of February 2010}

A strong outburst in multi-TeV gamma-rays from Mrk 421 was observed by VERITAS
telescopes  on 16th of February 2010 and follow up observations were
carried out by HESS telescopes from 17th to 20th of February a total
of 6.5 h. These data were taken in 11 runs with each run $\sim 28$
minutes duration\cite{Abeysekara:2016qwu}. The HESS telescopes
observed this flare in the energy range $1.67\, {\rm TeV} (4.0\times 10^{26}\, {\rm Hz}) \le E_{\gamma} \le 20.95\, {\rm TeV} (5.0\times 10^{27}\, {\rm Hz})$. As we have no multiwavelength SED obtained around the time of the Feb 2010 TeV flare, we use the models describing the observed SED at an earlier and later epochs. 
The first one is the same i.e. lep-1, which is used for the interpretation of the April 2004 flare. The second leptonic SED is from the
multiwavelength observation of Mrk 421 during 
January to March 2013, undertaken by GASP-WEBT, {\it Swift}, NuSTAR
Fermi-LAT, MAGIC, VERITAS\cite{Balokovic:2015dnz} and fitted with one-zone leptonic model
where the bulk Lorentz factor $\Gamma=25$ is used (lep-2).

\begin{figure}
{\centering
\resizebox*{0.5\textwidth}{0.35\textheight}
{\includegraphics{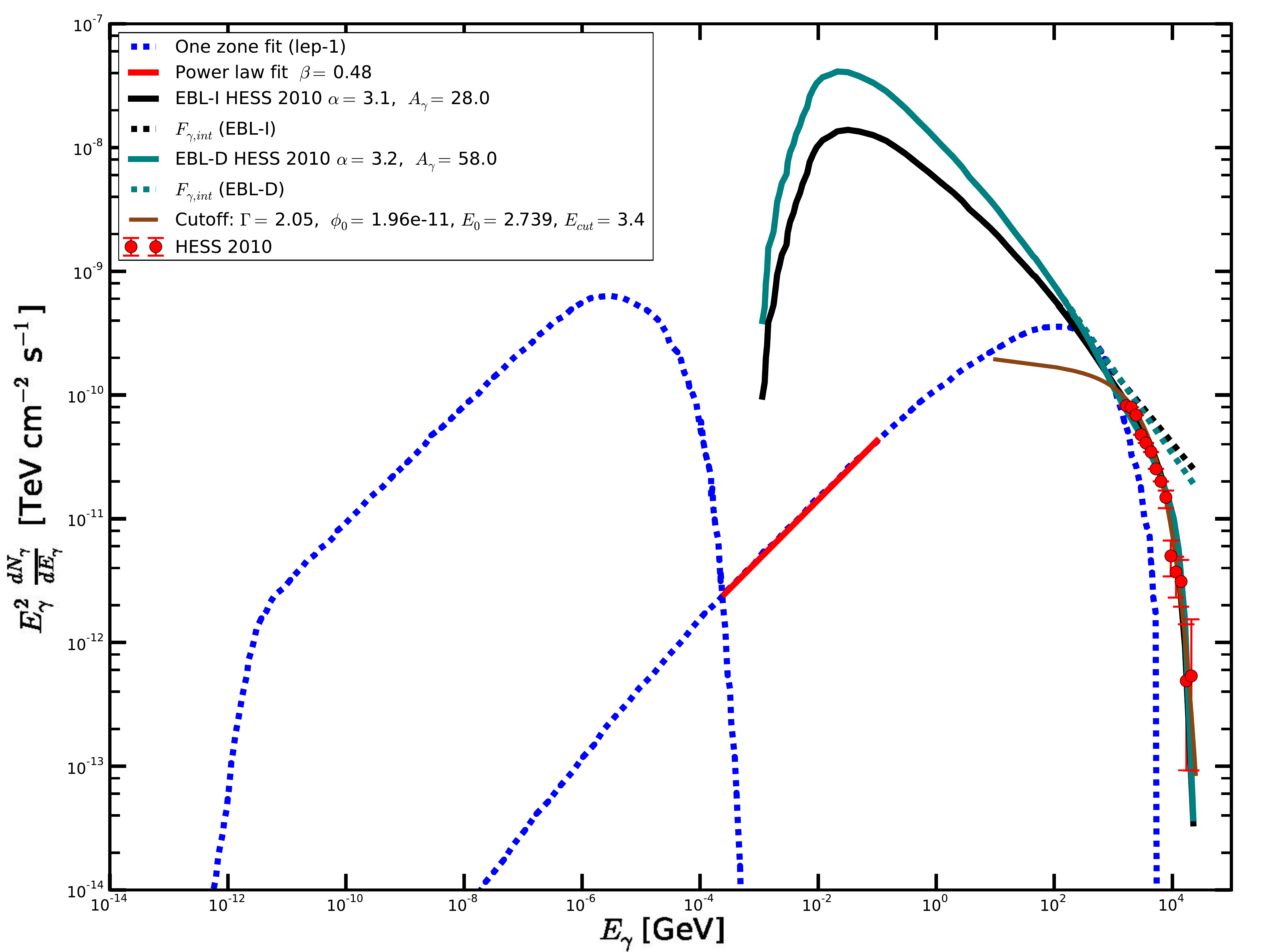}}
\par}
\caption{The SED of lep-1 is shown along with the power-law fit to the
  SSC tail region with $\beta=0.48$. The best fit to the flare
  data of 2010 using EBL-D and EBL-I are shown. A sharp increase
in the flux is observed at lower energy. We have also shown the
power-law with exponential cut-off fit for comparison\cite{Tluczykont:2011gs}.
\label{fig:figure4}
}
\end{figure}

\begin{table}[h]
\centering
\caption{A summary of the parameters used in the photohadronic fits
  for the observed data taken from the observations of Whipple in 2004
  and HESS in 2010. Here $\alpha$ and $A_{\gamma}$ are spectral index and
  normalization constant respectively.} 
\label{tab1}
\begin{tabular*}{\columnwidth}{@{\extracolsep{\fill}}llll@{}}
\hline
\multicolumn{1}{@{}l}{} & Lep-1 & Lep-2\\
\hline
${\cal D}$ (Doppler factor)& 14 & 25\\
$R^{\prime}_b$ (Blob radius) & $0.7 \times 10^{16} cm$ & $0.9 \times 10^{16} cm$& \\
$R^{\prime}_f$ (Inner blob radius)& $ \approx 3 \times 10^{15} cm$ & $ \approx 3 \times 10^{15} cm$\\
$B^{\prime}$ (Magnetic field)& 0.26 G& 0.17 G\\
\hline
EBL Model & $\alpha$,  $A_{\gamma}$ & $\alpha$,  $A_{\gamma}$ \\
\hline
EBL-I &2.7, 2.3 & 2.6, 2.4\\
EBL-D & 2.7, 2.5 & 2.6, 2.5 \\
\hline
\end{tabular*}
\end{table}

From the kinematical condition given in Eq. (\ref{KinCon}), the above range
of the observed gamma-ray corresponds to the  proton energy in the
range $16.7\, {\rm TeV} \le E_p \le 210\, {\rm TeV}$. Using lep-1, where
$\Gamma={\cal D}=14$, the seed
photon energy lies in the range $0.28\, {\rm MeV} (6.8\times 10^{19}\, {\rm Hz})
\le \epsilon_{\gamma} \le 3.53\, {\rm MeV} (8.5\times 10^{20}\, {\rm Hz})$ which
is again in the tail region of the SSC SED as shown in
Fig. \ref{fig:figure2}. Very good fit to the multi-TeV spectrum is obtained by using the
EBL-D and EBL-I. The best fit parameters are respectively $\alpha=3.1$ and
$A_{\gamma}=58.0$ for EBL-D and $\alpha=3.2$ and
$A_{\gamma}=28.0$ for EBL-I which correspond to very soft spectrum and
the intrinsic spectrum is also soft (between -0.68 to -0.58). Even
though these parameters fit well to the observed multi-TeV spectrum, 
in the low energy limit the spectrum shoots up very high as shown in
Fig. \ref{fig:figure4} which is certainly not observed by HESS
telescopes. So we can ignore these soft power-law fit to the observed
flare data and to look for $\alpha < 3$. This soft power-law problem arises
because $\beta=0.48$ is small. So we can use the leptonic model which
have $\beta > 0.48$ and will be able to get $\alpha < 3$.
The time averaged differential energy spectrum of this 
observation is also fitted with  a power-law with exponential cut-off
having four parameters\cite{Tluczykont:2011gs}.

\begin{figure}
{\centering
\resizebox*{0.5\textwidth}{0.35\textheight}
{\includegraphics{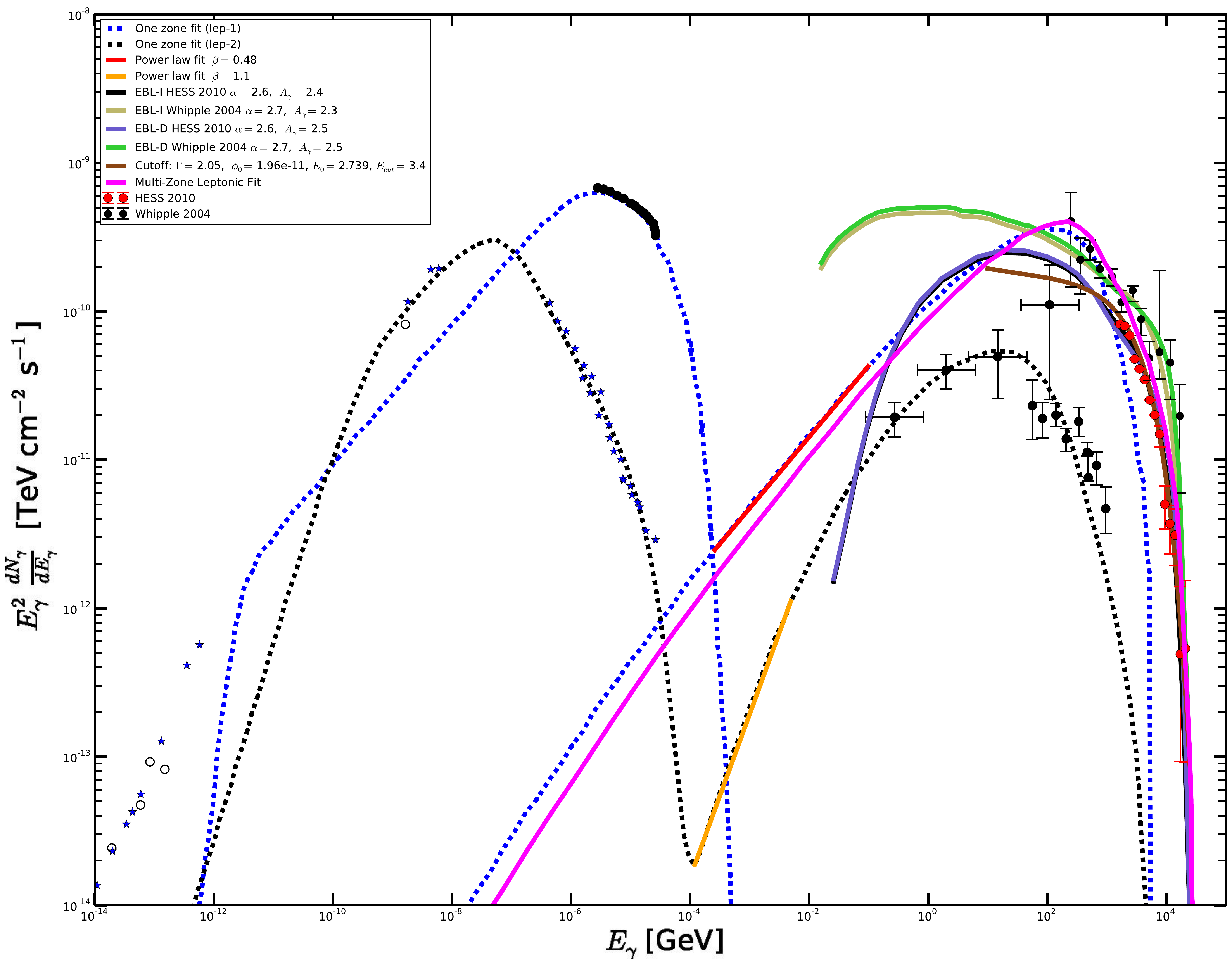}}
\par}
\caption{
The SED of lep-2\cite{Balokovic:2015dnz} is shown along with the power-law fit to the
  SSC tail region with $\beta=1.1$. The best fit to the flare
 of 2010 using EBL-D and EBL-I are also shown. For comparison, we have
 also shown the SED of lep-1, the power-law fit to the SSC tail region
 with $\beta=0.48$ and the best fit to the flare data of 2004 by
 Whipple telescope\cite{Blazejowski:2005ih}. The low energy observed
 data are taken from ref. \cite{Balokovic:2015dnz,Blazejowski:2005ih}.
\label{fig:figure5}
}
\end{figure}

\begin{figure}
{\centering
\resizebox*{0.5\textwidth}{0.35\textheight}
{\includegraphics{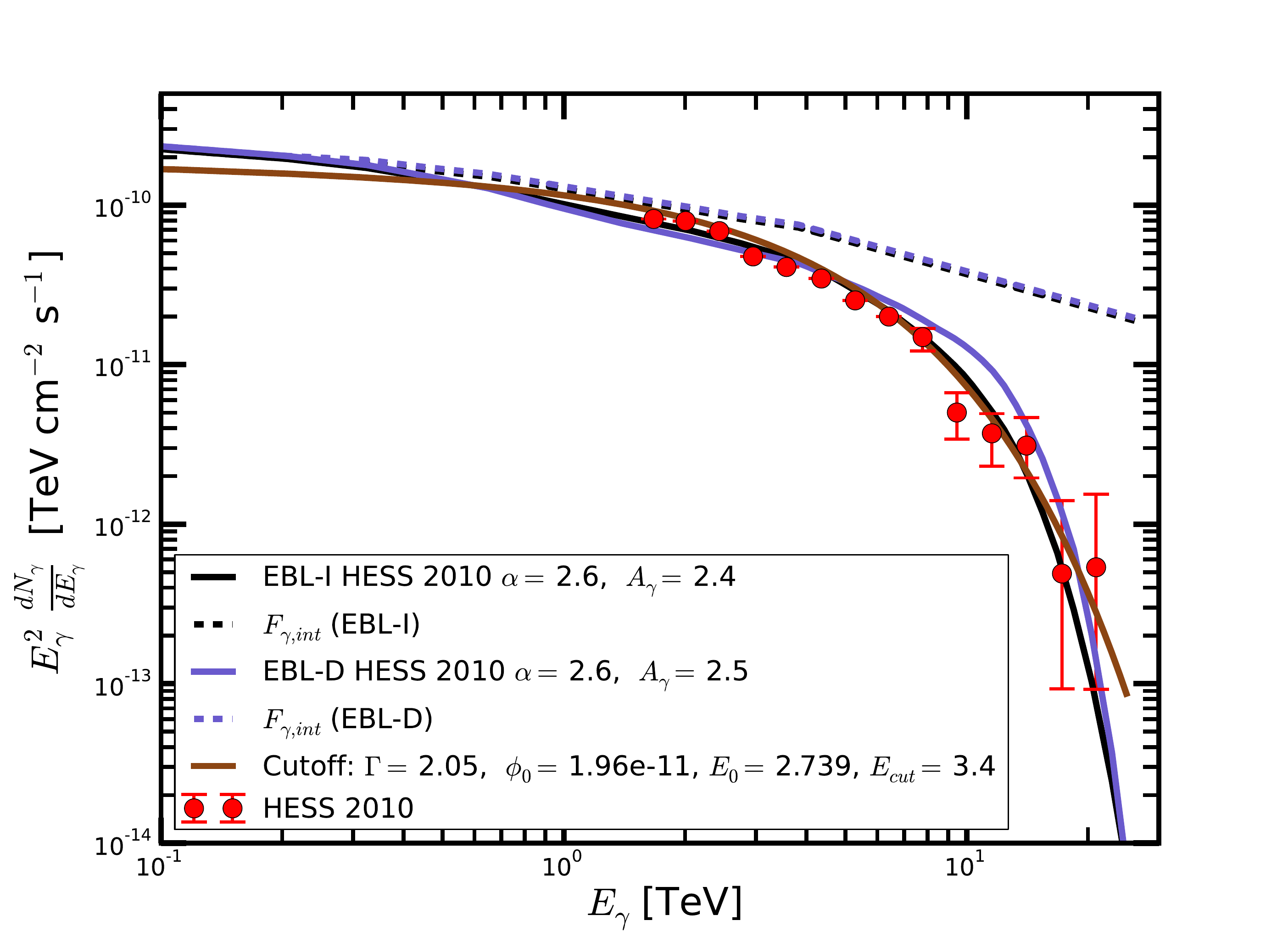}}
\par}
\caption{
Fit to the observed flux of 2010 flare by HESS  using photohadronic model
and EBL correction to it by EBL-D and EBL-I are shown. The corresponding intrinsic fluxes are
also given.
\label{fig:figure6}
}
\end{figure}

In the context of one-zone leptonic model lep-2 with $\Gamma={\cal D}=25$, the observed 
flare energy range $1.67\, {\rm TeV} (4.0\times 10^{26}\, {\rm Hz}) \le E_{\gamma} \le 20.95\, {\rm TeV} (5.0\times 10^{27}\, {\rm Hz})$ corresponds to the seed photon energy in the range 
$0.90\, {\rm MeV} (2.17\times 10^{20}\, {\rm Hz})$ $
\le \epsilon_{\gamma} \le 11.26\, {\rm MeV} (2.72\times 10^{21}\, {\rm Hz})$,
which is again in the tail region of the SSC SED as can be seen from
Fig. \ref{fig:figure5}. This is fitted with a 
power-law with $\beta=1.1$ and $\Phi_0=4.37\times 10^{-9}\, {\rm TeV} \,
{\rm cm^{-2}\, s^{-1}}$ (the pink line). 
Here again, we have used the EBL-D and EBL-I to fit the
2010 flare data in the photohadronic model which is also shown in
Fig. \ref{fig:figure5}. The best fit parameters are $\alpha=2.6$ and
$A_{\gamma}=2.5$ for EBL-D and $\alpha=2.6$ and
$A_{\gamma}=2.4$ for EBL-I respectively. We note that the observed
data is fitted well and at the same time, the flux decreases towards
low energy regime as expected with a peak flux of $F_{\gamma,peak}\sim 2.6\times
10^{-10}\, {\rm TeV} \, {\rm cm^{-2}\, s^{-1}}$ at 
$E_{\gamma}\sim 18\, {\rm GeV}$. Both the EBL-D and EBL-I corrections to the
photohadronic model give practically the same result. 
The parameters used in
the photohadronic model to fit the 2010 data are summarized in Table
\ref{tab1}.

Normally, the leptonic models fit the multiwavelength data
well. But as shown in Figs. \ref{fig:figure2} and  \ref{fig:figure5},
the leptonic models lep-1 and lep-2 do not fit to the radio (low
energy) data. 
It is possible that the low energy behavior of these models
have to be modified to accommodate these radio data.
The SSC photons are produced from the 
inverse Compton scattering of high energy electrons with the
synchrotron photons which obey the relation
$\epsilon_{SSC}\simeq 1.3
\gamma^2_e \epsilon_{syn}$, where $\epsilon_{SSC}$ is the SSC photon
energy, $\epsilon_{syn}$ is the synchrotron photon energy and
$\gamma_e$ is the electron Lorentz factor. From the energies of the
synchrotron peak and the SSC peak in lep-1, we obtain $\gamma_e\sim
2740$.  This implies that the 0.35 MeV SSC photon can be produced from
the boosting of the $\sim 3.5\times 10^{-11}$ GeV ($\sim 8.5\times
10^{12}$ Hz) synchrotron
photon which is in the infrared band. This clearly shows that low
energy photons in the radio band will not affect the prediction of
multi-TeV 
gamma-rays.

Even though lep-1 fits well to the multi-TeV data, in the low energy
regime, the flux increases drastically which is clearly shown in
Fig. \ref{fig:figure4}. This behavior is absent with the lep-2 fit.
It is to be noted that, lep-1 corresponds to the
observation during the year 2003-2004 and lep-2 is the recent one
of January 2013. So we believe that during each observation period,
the photon density distribution in the jet changes and this change in
the seed photon changes the spectral behavior of the observed
multi-TeV gamma-rays. This clearly shows that almost simultaneous observation
in multiwavelength is essential to fit the observed data.

In Fig. \ref{fig:figure6} we have plotted only the 2010 flare data along
with the EBL-D and EBL-I fits and their corresponding intrinsic
fluxes. We can observe a minor difference between EBL-D and EBL-I
predictions for $0.6\, {\rm TeV} \leq E_{\gamma} \le 20\, {\rm TeV}$.
The intrinsic flux is a power law with
$F_{\gamma,in}\propto E^{-0.7}_{\gamma}$. Comparison of
$F_{\gamma,in}$ of both 2004 and 2010 multi-TeV flaring shows that
different spectral shape of the observed events are solely due to the
diversity in the shape of the seed photon density distribution (particularly in the SSC
tail region) even though we have the same acceleration mechanism ($\alpha\simeq 2.6$) of protons in the
blazar jet. 

\section{Summary}

We used the photohadronic scenario
complemented by two EBL models (EBL-D and EBL-I) to interpret the multi-TeV
flares observed in April 2004 and February 2010. These EBL models
are equally good to explain the observed data which can also be seen from
the comparison of the attenuation factors in Fig. \ref{fig:figure1}. 
The photohadronic scenario with both the EBL models can fit   these
multi-TeV spectra very well and it is observed that the intrinsic
spectrum of each epoch is different even though the Fermi accelerated
high energy protons have almost the same spectral index $\alpha\simeq
2.6$. This difference in intrinsic spectra attributes to the change in
the spectral index of the seed photon in the SSC tail region of the
jet in different epochs of flaring. 
The same photohadronic models were earlier employed to explain the
multi-TeV emission from other HBLs\cite{Sahu:2016mww,Sahu:2018rqs}. 
We suggest that the same flaring mechanism may be acting in all HBLs.
The differences between the flaring behavior of HBL blazars and in one
blazar from flare to flare is determined by the changes in flux and
spectrum of the seed photons in the SSC region.
It is important to note that simultaneous or quasi-simultaneous
observation of the  tail region of the SSC SED  with TeV observation
would be most useful to constrain the photohadronic model.



We would like to thank the anonymous referee for valuable comments and suggestions. 
We thank D. Khangulyan, Yoshiyuki Inoue,
Susumu Inoue, Don Warren and Haoning He for many useful
discussions. S.S. is a Japan Society for the Promotion of Science
(JSPS) invitational fellow. A.R.L. is thankful to iTHEMS for financial
support and kind hospitality during his visit to RIKEN. This work is partially
supported by RIKEN, iTHEMS \& iTHES Program and also by Mitsubishi
Foundation. The work of S. S. is
partially supported by DGAPA-UNAM (Mexico) Project No. IN110815 and PASPA-DGAPA, UNAM.

\end{document}